\documentstyle[aps,pre,twocolumn,psfig,draft,floats]{revtex}

\newcommand{\eq}[1]{Eq.~(\ref{eq:#1})}
\newcommand{\fig}[1]{Fig.~\ref{fig:#1}}
\newcommand{\tbl}[1]{Table~\ref{tbl:#1}}
\newcommand{\exf}[2]{\mbox{$#1\!\times\! 10^{#2}$}}

\begin{document}

\wideabs{
\title{Dynamics of stable viscous displacement in porous media} 
\author{Eyvind Aker\thanks{Also at: Norwegian University of Science and Technology, N-7491 Trondheim, Norway}, Knut J\o rgen  M\aa l\o y}
\address{Department of Physics, University of Oslo, N-0316 Oslo, Norway}
\author{Alex Hansen}
\address{Department of Physics, Norwegian University of Science and 
Technology, N-7491 Trondheim, Norway}
\date{\today}
\maketitle

\begin{abstract}
We investigate the stabilization mechanisms of the invasion front in
two-dimensional drainage displacement in porous media by using a
network simulator. We focus on the process when the front stabilizes
due to the viscous forces in the liquids. We find that the capillary
pressure difference between two different points along the front
varies almost linearly as function of height separation in the
direction of the displacement. The numerical results support arguments
that differ from those suggested earlier for viscous
stabilization. Our arguments are based upon the observation that
nonwetting fluid flows in loopless strands (paths) and we conclude
that earlier suggested theories are not suitable to drainage when
nonwetting strands dominate the displacement process.  We also show
that the arguments might influence the scaling behavior between the
front width and the injection rate and compare some of our results to
experimental work.
\end{abstract}

}

\section{Introduction}
Immiscible displacement of one fluid by another fluid in porous media
has important applications in a wide range of different
technologies. Most often mentioned is hydrology and oil
recovery. From a theoretical point of view, the displacement process is
very complex and hard to describe in detail. Especially, much
attention has been paid to the rich variety of displacement
structures that is observed. The displacement structures are found to
depend strongly on the fluid properties like viscosity, interfacial
tension, fluid flow rate, and
wettability~\cite{Maloy85,Chen-Wilk85,Len88,Cieplak88}.

In drainage the primary process is the displacement of a wetting fluid
by a nonwetting fluid in porous media. Consider a two-dimensional (2D)
horizontal displacement of a less viscous fluid by a more viscous
fluid. At high injection rates the front developing between the
invading and defending fluid, is known to stabilize~\cite{Len88}.  In
contrast, at extremely low injection rate the invading fluid generates
a growing cluster similar to the cluster formed by invasion
percolation (IP)~\cite{Guyon78,Koplik82,Wilk83,Len85}.  The displacement is
now controlled solely by the capillary pressure, that is the pressure
difference between the two fluids across a meniscus.

\begin{figure}
\begin{center}
\mbox{\psfig{figure=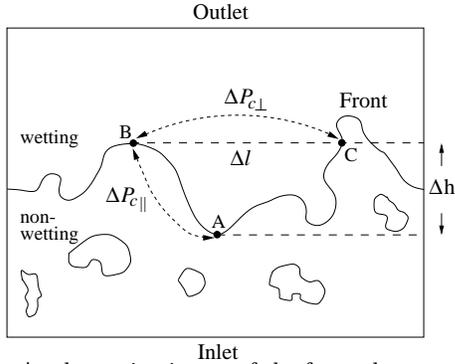,width=6cm}}
\caption{A schematic picture of the front that travels across the
system from the inlet to the outlet.  In the figure, $\Delta P_{c\|}$
is the capillary pressure difference between a meniscus at A and a
meniscus at B separated a vertical distance $\Delta h$. In the
orthogonal direction we calculate $\Delta P_{c\bot}$, that is the
capillary pressure difference between a meniscus at B and a meniscus
at C separated a horizontal distance $\Delta l$. $\Delta P_{nw}$ and
$\Delta P_w$ denote the viscous pressure drop going from A to B 
in the nonwetting and wetting phase, respectively.}
\label{fig:model}
\end{center}
\end{figure}

In this paper we address the question of how the invasion front
stabilizes when no gravity forces are present (2D horizontal
displacement). To do this, we have developed a network model that
properly simulates the dynamics of the capillary pressures due to the
menisci along the front as well as the viscous pressure buildup in the
fluids.  From the simulations we have calculated the capillary
pressure difference $\Delta P_{c\|}$ between menisci along the front
separated a distance $\Delta h$ in the direction of the displacement.
Also calculated, is the capillary pressure in the orthogonal direction
$\Delta P_{c\bot}$, that is the capillary pressure between menisci at
same height above the inlet but separated a horizontal distance
$\Delta l$ (see \fig{model}). Simulations show that assuming a power
law behavior $\Delta P_{c\|}\propto\Delta h^\kappa$, our best estimate
of the exponent for a wide range of injection rates and different
fluid viscosities is $\kappa = 1.0\pm 0.1$. This is a surprising
result because the viscous force field that stabilizes the front, is
non homogeneous due to trapping of wetting fluid behind the front and
to the fractal behavior of the front structure. 

We also presents arguments being supported by the numerical evidence
that $\kappa\simeq 1.0$. The arguments are based upon the observation
that nonwetting fluid displaces wetting fluid through loopless
strands (see \fig{40x60}). As a consequence, we find that existing
theories~\cite{Wilk86,Len89,Blunt92,Xu98} not considering this effect, are
not compatible with drainage when nonwetting strands dominate the
displacement process. We also conjecture that the result $\kappa\simeq 1.0$
may influence the scaling between the saturated front width $w_s$ and
the capillary number $C_a$. The capillary number is the ratio between
viscous and capillary forces and in the following $C_a\equiv
Q\mu_{nw}/\Sigma\gamma$. Here $Q$ is the injection rate, $\Sigma$ is
the cross section of the inlet, and $\mu_{nw}$ is the viscosity of the
nonwetting fluid.

The effect of gravity on the front when the fluids have different
densities has been thoroughly
discussed~\cite{Wilk84,Wilk86,Birov91,Birov92} and in slow drainage it
is found that gravity may stabilize the front. Gravity causes a
hydrostatic pressure gradient in the fluids and considering a heavy
nonwetting fluid below displacing vertically upwards a light wetting
fluid, this gradient will stabilize the front. The displacement
process corresponds exactly to IP with a stabilizing
gradient~\cite{Wilk86,Birov91,Sap88} and the saturated front width
$w_s$, has been shown to scale like $w_s\propto {B_o}^{-\nu/(1+\nu)}$.
Here $\nu$ is the correlation length exponent in percolation and $B_o$
is the bond number indicating the ratio between gravity and capillary
forces.

A similar consensus concerning the stabilization mechanisms when
viscous forces replace gravity forces has not yet been reached.  In
the literature the displacement has been related to
IP~\cite{Wilk86,Blunt92,Xu98}, however, the scenario is more
complicated than in the gravity case.  Gravity is a uniform force
acting on the whole system, while the viscous force is local and
fluctuates due to permeability variations and fluid trapping in the
porous medium. One standard approach is to separate the displacement
structure into two parts. One consisting of the frontal region, and
the other consisting of the static structure behind.  The frontal
region of extent $w_s$, is assumed to behave as the spanning cluster
in percolation. Consequently, it is assigned the permeability
$k\propto w_s^{-t/\nu}$, where $t$ is the conductivity exponent in
percolation.  By applying Darcy's law and assuming that the stabilized
front reaches a traveling-wave state according to Buckley-Leverett
displacement~\cite{Buckley42}, the scaling of the front width is found
to behave as $w_s\propto {C_a}^{-\alpha}$. In the literature there
exists two slightly different expression for $\alpha$.  In 3D
Wilkinson~\cite{Wilk86} found $\alpha=\nu/(1+t-\beta+\nu)$ where
trapping of wetting fluid is assumed to be less important. Here
$\beta$ is the order parameter exponent in percolation. Later, Blunt
{\em et al.}~\cite{Blunt92} suggested in 3D that
$\alpha=\nu/(1+t+\nu)$ which is identical to the result of
Lenormand~\cite{Len89} discussing limits of fractal patterns between
capillary fingering and stable displacement in 2D porous media. In
Appendix~\ref{app:alex} we present a simple method giving
$\alpha=\nu/(1+t-\beta+\nu)$ by applying percolation concepts on the
frontal region when not considering that nonwetting fluid flows in strands.

Recently, Xu {\em et al.}~\cite{Xu98} used Wilkinson's arguments and
deduced a scaling relation for the viscous pressure drops in the
frontal region. They proposed that the nonwetting pressure drop
$\Delta P_{nw}$ in the front (see \fig{model}) should scale as $\Delta
P_{nw}\propto \Delta h^{t/\nu+d_{\text{E}}-1-\beta/\nu}$ over a
distance $\Delta h$ in the direction of the displacement. Here,
$d_{\text{E}}$ is the Euclidean dimension of the space in which the 
front is embedded (in our case $d_{\text{E}}=2$) and $\Delta h$ is assumed to
be sufficiently large for scaling to be acceptable and less than
$w_s$. They also argued that the pressure drop in the wetting phase
$\Delta P_{w}$, must be linearly dependent on $\Delta h$, since the
displaced phase is compact. In~\cite{Blunt92} Blunt {\em et al.}  also
suggested a scaling relation for $\Delta P_{nw}$, however, in 3D they
found $\Delta P_{nw}\propto\Delta h^{t/\nu+1}$. This deviates from the
result of Xu {\em et al.} when $d_{\text{E}}=3$. 

The paper is organized as follows. In Sec.~\ref{sec:model} we describe
the network model used in the simulations. Sec.~\ref{sec:result}
contains the simulation results of $\Delta P_{c\|}$ and $\Delta
P_{c\bot}$, supporting the arguments we present in
Sec.~\ref{sec:trap}. In Sec.~\ref{sec:expr} we compare our findings to
some experimental data and the conclusions are drawn in
Sec.~\ref{sec:concl}. At the end we have put Appendix~\ref{app:alex}
where we deduce the scaling relation between $w_s$ and $C_a$ using the
ideas in~\cite{Wilk86} when not considering the effect of nonwetting
fluid flowing in strands.

\section{Network Model}\label{sec:model}
The network model has been presented elsewhere~\cite{Aker98-1,Aker98-2} and
therefore only its main features will be given here.

In the simulations we have constructed the porous medium in two
different ways. In the first way the porous medium is represented by a
square lattice of tubes oriented at $45^\circ$. The tubes are
cylindrical with length $d$. Each tube between the $i$th and the $j$th
node in the lattice is assigned an average radius $r_{ij}$ which is
chosen at random in the interval $[\lambda_1 d,\lambda_2 d]$, where
$0\le\lambda_1 < \lambda_2\le 1$.  The randomness of the radii
represents the disorder in the system. In the following this system
will be referred to as the random radii lattice.

In the second way the porous medium is constructed upon a square
lattice inclined $45^\circ$ where the distance between each
intersection in the lattice is of unit length.  Around each
intersection we draw a circle of radius $\lambda$. To avoid
overlapping circles the given $\lambda$ must be in the interval
$0\le\lambda<1/2$. A node is placed at random inside each of the
circles and the nodes inside the nearest neighbor circles are
connected by cylindrical tubes. Thus, as for the random radii
lattice, four tubes meet at each node.  We let $d_{ij}$ denote the
length of the tube between the $i$th and $j$th node, and the
corresponding radius $r_{ij}$ is defined as
$r_{ij}=d_{ij}/2\alpha$. Here $\alpha$ is the aspect ratio between the tube
length and the radius. In the simulations $\alpha = 1.25$, hence, the
tubes are $25\%$ longer than they are wide.  In this lattice the
position of the nodes represent the disorder in the system, and
therefore we will refer to it as the random node lattice.

While every pair of nearest neighbor nodes
are separated an equal distance in the random radii lattice,
the distance between two nearest neighbor nodes vary in the random
node lattice. Especially, the shortest length scale, that is the
minimum distance between two neighboring nodes, is less in the random
node lattice. Consequently, we are able to generate more narrow fronts
at higher injection rates in the random node lattice, than what is
possible with the random radii lattice. Therefore the random
node lattice is preferred at high injection rates where a flat front
is generated.

In both lattices the tubes represent the volume of both pores and
throats, and there is no volume assigned to the nodes. The liquids flow
from the bottom to the top of the lattice, and we implement periodic
boundary conditions in the horizontal direction. The pressure
difference between the bottom row and the top row defines the pressure
across the lattice. Initially, the system is filled with a wetting
fluid with viscosity $\mu_{w}$. The injected fluid is nonwetting and
has viscosity $\mu_{nw}\ge\mu_{w}$. The viscosity ratio $M$, is defined
as $M\equiv\mu_{nw}/\mu_{w}$. 

The capillary pressure $p_c$ between the nonwetting and wetting fluid
in a tube is given by Young-Laplace law
\begin{equation}
p_c=\gamma\left(\frac{1}{R_1}+\frac{1}{R_2}\right),
\label{eq:laplace}
\end{equation}
where $R_1$ and $R_2$ are the principal radii of curvature of the
interface (a meniscus) and $\gamma$ is the interfacial tension. In a
cylindrical tube of radius $r$ where $R_1=R_2$,
\eq{laplace} reduces to
$p_c=(2\gamma/r)\cos\theta$. Here $\theta$ denotes the wetting angle
between the nonwetting and wetting phases, and in drainage $\theta$ is
in the interval $(0,\pi/2)$.

In the network model we treat the tubes as if they were hourglass
shaped with effective radii following a smooth function.  Hence, we
let the capillary pressure become a function of the meniscus position
in the tube and assume the Young-Laplace law~(\ref{eq:laplace}) takes
the form
\begin{equation}
p_c=\frac{2\gamma }{r}\left[ 1-\cos (2\pi\mbox{\small $\frac{x}{d}$})\right].
\label{eq:pcvary}
\end{equation}
Here $0\le x\le d$ is the position of the meniscus in the tube where
$d$ is the tube length. We assume perfect wetting, i.e.\ $\theta=0$.

By letting $p_c$ vary according to~(\ref{eq:pcvary}), we include the
effect of burst dynamics into the model~\cite{Aker98-1}. This is
particularly seen at low injection rates where the invasion of
nonwetting fluid takes place in bursts accompanied by sudden negative
jumps in the pressure (Haines jumps)~\cite{Hain30,Maloy92,Maloy96}.  The
detailed modelling of the capillary pressure costs computation time.
However, it is necessary in order to properly simulate the pressure
behavior along the front.

The volume flux $q_{ij}$ through a tube from the $i$th to the $j$th
node is found from the Washburn equation for capillary flow~\cite {Wash21}
\begin{equation}
q_{ij}=-\frac{\sigma_{ij} k_{ij}}{\mu_{ij}}\frac{1}{d_{ij}}(\Delta p_{ij}-p_{c,ij}).
\label{eq:tubeflow}
\end{equation}
Here $k_{ij}$ is the permeability of the tube ($r_{ij}^2/8$) and
$\sigma_{ij}$ is the cross section ($\pi r_{ij}^2$) of the tube.  $\mu_{ij}$ is
the effective viscosity given by the sum of the volume fractions of
each fluid inside the tube multiplied by their respective
viscosities. The pressure drop across the tube is $\Delta
p_{ij}=p_j-p_i$, where $p_i$ and $p_j$ is the nodal pressures at node
$i$ and $j$ respectively.  The capillary pressure $p_{c,ij}$ is the sum
of the capillary pressures of the menisci (given by \eq{pcvary})
inside the tube. A tube
partially filled with both liquids, is allowed to contain either one or
two menisci. For a tube without menisci $p_{c,ij}=0$, and \eq{tubeflow}
reduces to that describing Hagen-Poiseuille flow with $\mu_{ij}=\mu_1$
or $\mu_2$.

We assume conservation of volume flux at each node giving
\begin{equation}
\sum_j q_{ij}=0.
\label{eq:Kirch}
\end{equation}
The summation on $j$ runs over the nearest neighbor 
nodes to the $i$th node while $i$ runs over all nodes that do not 
belong to the top or bottom rows, that is, the internal nodes. 

Eqs.~(\ref{eq:tubeflow}) and~(\ref{eq:Kirch}) constitute a set of
linear equations which are to be solved for the nodal pressures $p_i$,
with the constraint that the pressures at the nodes belonging to the
upper and lower rows are kept fixed.  The set of equations is solved
by using the Conjugate Gradient method~\cite{Bat88}. 

During every simulation we held the injection rate $Q$ fixed and
calculate a time dependent pressure $\Delta P$ across the system. See
Refs.~\cite{Aker98-1,Aker98-2} for details on how $\Delta P$ and
the corresponding $p_i$'s are found. 

Having found the  $p_i$'s we calculate the volume
fluxes, $q_{ij}$, through every tube in the network, using
\eq{tubeflow}. According to the $q_{ij}$'s we define a time step
$\Delta t$, such every meniscus is allowed to travel at most a maximum
step length $\Delta x_{\text{max}}$, during that time step. The
menisci are then moved a distance $(q_{ij}/\sigma_{ij})\Delta t$ and
the pressure $\Delta P$ and the time lapse are recorded, before the
$p_i$'s are solved for the new fluid configuration.  Menisci that are
moved out of a tube during a time step are spread into neighbor
tubes. For details about how the menisci is moved into neighbor tubes
see Refs.~\cite{Aker98-1,Aker98-2}.

Numerical simulations show that $\Delta x_{\text{max}}$ must be of
order $0.1$ to calculate the variation in the capillary pressure when
a meniscus travel through a tube. In all simulations presented here
$\Delta x_{\text{max}} = 0.1$, resulting in at least ten time steps to
invade one tube with nonwetting fluid.  This causes the computation
time to increase dramatically and one displacement simulation on
lattices of sizes presented in this paper takes typically between
3--15 hours on a 400 MHz Pentium II processor.

\section{Simulations}\label{sec:result}

We have run drainage simulations at different injection rates and
fluid viscosities to study the capillary pressure variations along the
invasion front. Due to the huge computational effort that is
necessary, the simulations have been limited to lattices of size
$25\times 35$ and $40\times 60$ nodes (Sec.~\ref{sec:pc}).  We have
also run some simulations where the lattice initially was filled with
nonwetting and wetting fluid according to patterns which were
generated by an IP algorithm (Sec.~\ref{sec:ip}).  In this way, we
were able to study the capillary pressure along invasion fronts on
lattices of $200\times 300$ nodes.

In every simulation, $\Delta P_{c\|}$, $\Delta P_{c\bot}$, and the
front width between the invading and the defending fluid, was
recorded.  The front was detected by running a Hoshen-Kopelman
algorithm~\cite{Stauf92} on the lattice and recognized as the set of
tubes that contain a front meniscus between the nonwetting and wetting
phase, that is the front tubes. The front width $w$ is defined as the
standard deviation of the vertical distances between front tubes and
the average position of the front. Let $h_i$ denote the vertical
distances of the front tubes above the inlet, where $i=1,...,n_f$ and
$n_f$ is the total number of front tubes. Then at a particular time,
we calculate $w=[(1/n_f)\sum_i(h_i-h)^2]^{1/2}$, where $h$ is the
average of the $h_i$'s.

$\Delta P_{c\|}$ and $\Delta P_{c\bot}$ is calculated as follows.
Consider two front menisci denoted by $m$ and $n$ with height $h_m$
and $h_l$ above the inlet (bottom row) at a distance $l_m$ and $l_n$
from the left boundary of the lattice. Assume that $h_m > h_n$, then
we calculate the difference $\Delta {P_c}^{mn}(\Delta h,\Delta l) =
p_c^n-p_c^m$ where $\Delta h = h_m-h_n$ and $\Delta l = |l_m-l_n|$. If
instead $h_n > h_m$, we compute $\Delta {P_c}^{nm}(\Delta h,\Delta l)
= p_c^m-p_c^n$ where $\Delta h = h_n-h_m$. We only consider the front
tubes containing one meniscus and where the nonwetting fluid invades
the tube from below. Note also, that we always take the capillary
pressure of the meniscus closest to the inlet minus the capillary
pressure of the meniscus closest to the outlet.  From above, we define
$\Delta P_{c\|}$ as function of $\Delta h$ as the average of $\Delta
{P_c}^{mn}$ over all pairs $mn$ separated a distance $\Delta h$ but
different $\Delta l$, i.e $\Delta P_{c\|}=\langle\Delta
{P_c}^{mn}(\Delta h=\mbox{const.}\, , \Delta l)\rangle$.

The capillary pressure difference in the orthogonal direction, $\Delta
P_{c\bot}$, (parallel to the inlet) as function of $\Delta l$ is
defined as the average of $|\Delta {P_c}^{mn}|$ over all pairs $mn$
with equal height ($\Delta h=0$) above the inlet when $\Delta l$ is held
constant. Thus, in the above notation $\Delta P_{c\bot}= \langle|\Delta
{P_c}^{mn}(0,\Delta l=\mbox{const.})|\rangle$.

The simulations were performed with parameters as close as possible to
experiments performed in~\cite{Frette97}. In the random radii lattice
we set the length $d$, of all tubes equal to $1\ \mbox{mm}$ and
the radii $r$ of the tubes were randomly chosen in the interval
$0.05d\leq r\leq d$.  In the lattices with random nodes we chose the
positions of the nodes such that the length of the tubes were inside
the interval $0.2\le d\le 1.8\ \mbox{mm}$. This gave us the radii of
the tubes, defined by $r=d/2\alpha$, where $\alpha=1.25$.  For both
types of lattices the interfacial tension was set to $\gamma=30\
\mbox{dyn/cm}$, and the fluid viscosities were 0.10 P,
0.50 P or 10 P.

\subsection{Capillary pressure behavior\label{sec:pc}}

We have performed two series of simulations with viscosity ratio
$M=100$ and one series of viscosity matched fluids, $M=1$. In all series
the capillary number $C_a$, was systematically varied by changing the
injection rate $Q$. Tables~\ref{tbl:m100-1}, \ref{tbl:m100-2},
and~\ref{tbl:m1} list $Q$, $C_a$ and the type of
lattice (random radii or random nodes) used in the different series.
Also shown are the calculated front width $w_s$, and the number of
different runs we did at each $Q$ to obtain reliable average
quantities.

 \begin{table}
\caption{Simulations performed on the random radii lattice of size
$25\times 35$ nodes and $M=100$ ($\mu_{nw}=10$ P, $\mu_w=0.10$ P).
The table contains the number of runs at each $Q$ and $C_a$ and the
calculated $w_s$.}
\label{tbl:m100-1}
\begin{tabular}{cdcc}
Runs &        $Q$                 & $C_a$ & $w_s$       \\ 
     & ($\mbox{cm}^3/\mbox{min}$) &       &              \\ \hline 
30   &         0.050   &  $\exf{3.7}{-4}$ & $5.5\pm 0.5$ \\
30   &         0.10    &  $\exf{7.3}{-4}$ & $4.3\pm 0.4$ \\
30   &         0.20    &  $\exf{1.5}{-3}$ & $3.7\pm 0.4$ \\
30   &         0.50    &  $\exf{3.7}{-3}$ & $3.0\pm 0.3$ \\
30   &         0.80    &  $\exf{5.8}{-3}$ & $2.5\pm 0.3$ \\
30   &         1.5     &  $\exf{1.1}{-2}$ & $2.4\pm 0.2$ 
\end{tabular}
\end{table}

\begin{table}
\caption{Simulations performed on the random node lattice of size
$25\times 35$ nodes and $M=100$ ($\mu_{nw}=10$ P, $\mu_w=0.10$ P). The
table contains the number of runs at each $Q$ and $C_a$, and the
calculated $w_s$.}
\label{tbl:m100-2}
\begin{tabular}{cdcc}
Runs &        $Q$                 & $C_a$ & $w_s$       \\ 
     & ($\mbox{cm}^3/\mbox{min}$) &       &             \\ \hline 
10   &         0.010   &  $\exf{1.0}{-4}$ & $4.3\pm 0.6$ \\
20   &         0.030   &  $\exf{3.1}{-4}$ & $2.9\pm 0.3$ \\
20   &         0.050   &  $\exf{5.2}{-4}$ & $2.5\pm 0.2$ \\
20   &         0.10    &  $\exf{1.0}{-3}$ & $2.1\pm 0.2$ \\
20   &         0.30    &  $\exf{3.1}{-3}$ & $1.4\pm 0.1$ \\
15   &         0.50    &  $\exf{5.2}{-3}$ & $1.2\pm 0.1$ \\
15   &         1.0     &  $\exf{1.0}{-2}$ & $0.9\pm 0.1$ \\
10   &         2.0     &  $\exf{2.1}{-2}$ & $0.8\pm 0.1$ \\
10   &         4.0     &  $\exf{4.2}{-2}$ & $0.8\pm 0.1$ 
\end{tabular}
\end{table}

\begin{table}
\caption{Simulations performed on the random node lattice of size
$40\times 60$ nodes and $M=1$ ($\mu_{nw}=\mu_w=0.50$ P). The table
contains the number of runs at each $Q$ and $C_a$ and the calculated
$w_s$.}
\label{tbl:m1}
\begin{tabular}{cdcc}
Runs &            $Q$             & $C_a$            & $w_s$ \\
     & ($\mbox{cm}^3/\mbox{min}$) &                  &              \\ \hline 
 10  &          0.050             &  $\exf{1.6}{-5}$ & $7.5\pm 1.5$ \\
 10  &          0.10              &  $\exf{3.2}{-5}$ & $6.9\pm 1.2$ \\
 15  &          0.30              &  $\exf{9.7}{-5}$ & $5.2\pm 0.5$ \\
 15  &          0.60              &  $\exf{1.9}{-4}$ & $4.4\pm 0.5$ \\
 20  &          1.2               &  $\exf{3.9}{-4}$ & $3.8\pm 0.5$ \\
 20  &          2.4               &  $\exf{7.8}{-4}$ & $3.0\pm 0.2$ \\
 20  &          4.8               &  $\exf{1.6}{-3}$ & $2.4\pm 0.2$ 
\end{tabular}
\end{table}

\begin{figure}
\begin{center}
\mbox{\psfig{figure=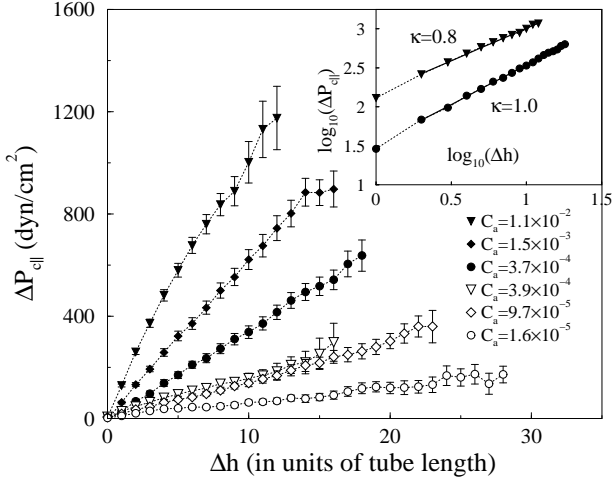,width=8cm}}
\caption{$\Delta P_{c\|}$ as function of $\Delta h$ for some $C_a$
with $M=100$ (Table~\protect\ref{tbl:m100-1}) and $M=1$
(Table~\protect\ref{tbl:m1}). $\Delta P_{c\|}$ is the average of the
different runs performed at each $C_a$, and the error bars denote the
standard error of the mean. {\em Inset:} $\log_{10}(\Delta P_{c\|})$
as function of $\log_{10}(\Delta h)$ for $C_a=\exf{1.1}{-2}$ and
$C_a=\exf{3.7}{-4}$ with $M=100$.  The solid lines were fitted to the
curves and their slopes are given by $\kappa$.}
\label{fig:pcdiff}
\end{center}
\end{figure}

\fig{pcdiff} shows the calculated capillary pressure difference
$\Delta P_{c\|}$, in the direction of the displacement as function of
height separation $\Delta h$. We have plotted the result for some of
the simulations performed on the random radii lattice of $25\times 35$
nodes with $M=100$ (filled symbols) and for some of the random node
lattice of $40\times 60$ nodes with $M=1$ (open symbols). In the inset
of \fig{pcdiff} the results for highest and lowest $C_a$ with $M=100$,
are plotted in a logarithmic plot and fitted to straight lines.
Assuming a power law behavior, we find that at $C_a=\exf{3.7}{-4}$ and
$M=100$, $\Delta P_{c\|}\propto\Delta h^{\kappa}$ and $\kappa=1.0$.
The exponent $\kappa$ seems to decrease systematically with increasing
injection rate, and at $C_a=\exf{1.1}{-2}$ and $M=100$ our best
estimate is $\kappa = 0.8$. Similar results was found from the
simulations performed with viscosity matched fluids ($M=1$). The data
points corresponding to $\Delta h\le 1$ tube length, is omitted in the
calculations of the exponent in \fig{pcdiff}. At short distances we
expect uncertainties in the result because of the finite length of the
tubes in the lattice.

In \fig{pcdiff} we observe that $\Delta P_{c\|}$ increases more
rapidly as function of $\Delta h$ at high injection rates compared to
the results at low injection rates. In the plot the effect is most
significant when $M=100$. At extremely low injection rate we expect
$\Delta P_{c\|}$ in \fig{pcdiff} to approach zero and become
independent of $\Delta h$. In this limit the capillary pressure of the
menisci along the front are almost equal (capillary equilibrium). As
seen from \fig{pcdiff}, we have not performed simulations with
that low injection rate. Instead the lowest $C_a$ for $M=100$ and
$M=1$, corresponds to the injection rate where no clear stabilization
of the front was found due to the finite size of the system.

\begin{figure}
\begin{center}
\mbox{\psfig{figure=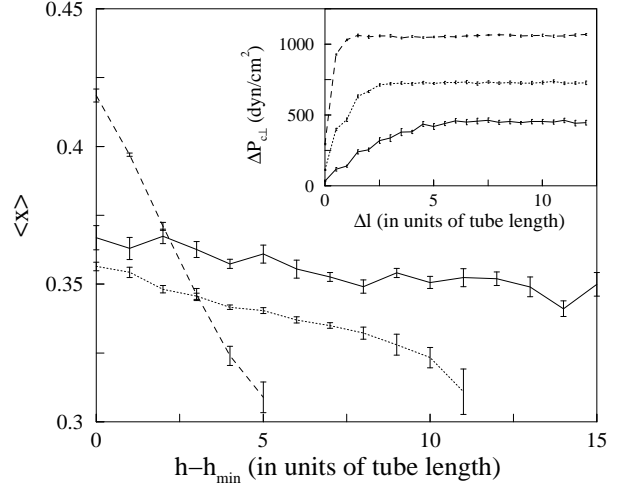,width=8cm}}
\caption{The average position $\langle x\rangle$, of the front menisci
inside the tubes as function of the menisci's height $h$ relative to
the of bottom height of the front $h_{\text{min}}$. The plot shows
result from simulations in \protect\tbl{m100-2} at
$C_a=\protect\exf{1.0}{-4}$ (solid line), \protect\exf{1.0}{-3}
(dotted line), and \protect\exf{1.0}{-2} (dashed line). {\em Inset:}
The corresponding $\Delta P_{c\bot}$ as function of $\Delta l$.  The
lattice size was $25\times 35$, giving a maximum horizontal distance
$\Delta l=12.5$ due to the periodic boundary conditions in the
horizontal direction.  The error bars denote the standard error in the
mean.}
\label{fig:xpos-pch}
\end{center}
\end{figure}

At higher injection rates the viscous gradient stabilizes the
front. The gradient results the capillary pressure of the menisci
closest to the inlet to exceed the capillary pressure of the menisci
further down the stream. This is indicated in \fig{xpos-pch},
showing the average position $\langle x \rangle$ of the front menisci
inside the tubes as function of their vertical height $h$, relative to
the bottom height of the front $h_{\text{min}}$. $\langle x \rangle$
is plotted for high, intermediate, and low $C_a$ for the simulations
listed in \tbl{m100-2}. From the figure we observe that at high
$C_a=\exf{1.0}{-2}$ (dashed line), the menisci near $h_{\text{min}}$
is placed closer to the middle of the tube compared to the menisci
ahead.  Consequently, the capillary pressure of the menisci near
$h_{\text{min}}$ will more likely be larger than the capillary
pressure of the menisci away from $h_{\text{min}}$ and therefore tubes
near $h_{\text{min}}$ will more easily be invaded.  This will
eventually stabilize the front. Remember that the tubes are hourglass
shaped and most narrow at $x=0.5$ (see
\eq{pcvary}).  At low injection rate, $C_a=\exf{1.0}{-4}$ (solid
line), we approach the regime of capillary equilibrium giving almost
no difference in $\langle x \rangle$ as function of
$h-h_{\text{min}}$.

For the three $C_a$'s in \fig{xpos-pch} we have also calculated the
capillary pressure difference in the orthogonal direction, $\Delta
P_{c\bot}$, as function of horizontal distance, $\Delta l$. The result
is shown in the inset of \fig{xpos-pch}. Here we interpret $\Delta
P_{c\bot}$ as the horizontal correlations in the capillary pressure
between menisci at same height.  Recall that $\Delta P_{c\bot}$
contains terms like $|p_c^m-p_c^n|=\sqrt{(p_c^m-p_c^n)^2}$, where
$p_c^m$ and $p_c^n$ denote the capillary pressure of two front menisci
$m$ and $n$, respectively. From the inset of \fig{xpos-pch} we see
that at low $C_a=\exf{1.0}{-4}$ (solid line) the capillary pressure of
two menisci at same height and a distance $\Delta l\lesssim 7$ apart,
are correlated to each other because $\Delta P_{c\bot}$ as not yet
reached the constant plateau ($\Delta l>7$) where the capillary
pressures becomes uncorrelated. At short distances $\Delta P_{c\bot}$
approaches zero, indicating that neighboring menisci have equal
capillary pressures. At high $C_a=\exf{1.0}{-2}$ (dashed line), we
observe that the correlations are very short.  Already for $\Delta
l>1$, $\Delta P_{c\bot}$ reaches the plateau and the capillary
pressures of the menisci do no longer interfere. Thus, if we consider
a narrow and a wide tube at same height, the viscous forces are strong
enough to push the nonwetting fluid through both the narrow and the
wide tube simultaneously.  As a result nonwetting fluid will invade
simultaneously everywhere along the front.  Similar behavior is
observed in the other simulations listed in Tables~\ref{tbl:m100-1}
and~\ref{tbl:m1} at high $C_a$

\subsection{Effect of viscosity ratio on the capillary 
pressure\label{sec:m-depend}}

\begin{figure}
\begin{center}
\mbox{\psfig{figure=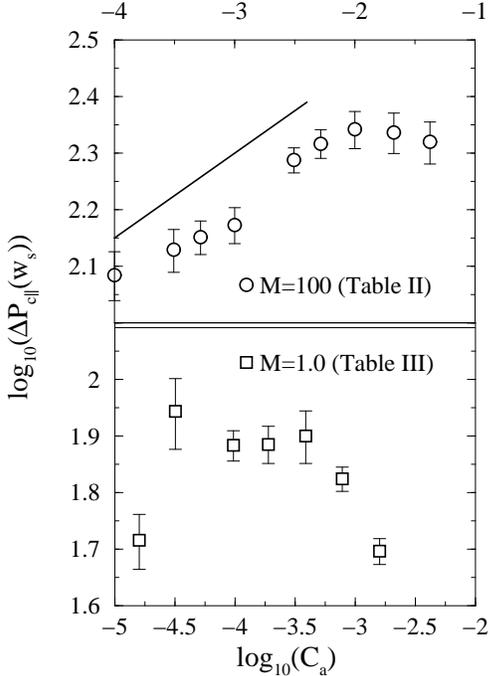,width=6.5cm}}
\caption{$\log_{10}(\Delta P_{c\|}(w_s))$ as function of
$\log_{10}(C_a)$ for the simulations performed on the random node
lattice with $M=100$ (top) and $M=1$ (bottom).  The slope of the solid
line in the upper figure is 0.15.  The error bars denote the standard
error in the mean.}
\label{fig:pcdmax}
\end{center}
\end{figure}

\fig{pcdmax} shows a log-log plot of $\Delta P_{c\|}$ taken at $\Delta
h=w_s$, as function of $C_a$ for the simulations performed on the
random node lattice with $M=100$ (\tbl{m100-2}) and $M=1$ (\tbl{m1}).
In the following $\Delta P_{c\|}$ at $w_s$ is denoted as $\Delta
P_{c\|}(w_s)$. If we ignore the effect of nonwetting strands and use
the result presented in Appendix~\ref{app:alex} on our problem, we
have that $\Delta P_{c\|}(w_s)\propto C_a{w_s}^\kappa$ by setting
$\Delta h=w_s$ in \eq{Pc}. Here $w_s\propto {C_a}^{-\alpha}$ where
$\alpha=\nu/(1+t-\beta+\nu)$ and $\kappa = t/\nu+1-\beta/\nu$
according to Appendix~\ref{app:alex}. By combining the two power laws
we obtain $\Delta P_{c\|}(w_s)\propto {C_a}^{1/(1+t-\beta+\nu)}$
giving in 2D, $\Delta P_{c\|}(w_s)\propto {C_a}^{0.29}$.

If we assume a power law behavior between $\Delta P_{c\|}(w_s)$ and
$C_a$, our best result for the exponent is $0.15\pm
0.05$ when $M=100$ in \fig{pcdmax}.  Note that there seems to be an
upper cut off at $C_a\gtrsim\exf{1.0}{-2}$ where $\Delta P_{c\|}(w_s)$
stops growing.  At $C_a\gtrsim\exf{1.0}{-2}$ the front is typically
flat and we approach the minimum width due to the finite length of the
tubes (see \tbl{m100-2}). In this limit we expect a cross over to
another type of behavior.

If it is difficult to confirm any power law when $M=100$, the
result of $M=1$ in \fig{pcdmax} does not show any scaling behavior.
Already for $C_a\gtrsim\exf{1}{-4}$, $\Delta P_{c\|}(w_s)$
reaches a plateau or even decreases. To explain the different behavior
of $\Delta P_{c\|}(w_s)$ when $M=1$ and $100$, we first look at the
strength of the capillary pressure drop across the front and
second we compare that to the magnitude of $\Delta P_{c\bot}$ as
function of $C_a$.

\begin{figure}
\begin{center}
\mbox{\psfig{figure=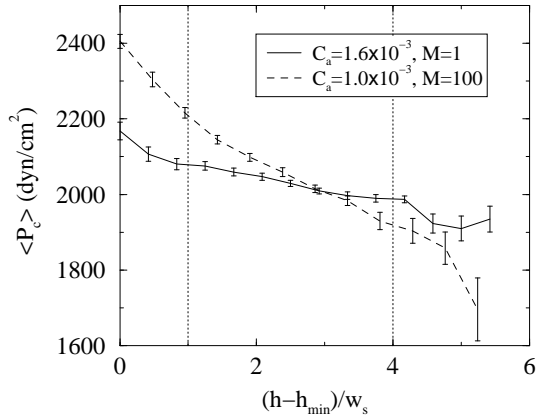,width=7cm}}
\caption{$\langle P_c\rangle$ in the frontal region as function of the
relative height from the bottom of the front.  The height distance is
normalized by dividing with the saturated front width $w_s$. The
vertical dashed lines indicate the region where $\langle P_c\rangle$
is approximately linear. The error bars denote the standard error of
the mean.}
\label{fig:pcgrad}
\end{center}
\end{figure}

To study the capillary pressure drop we have calculated the average
capillary pressure $\langle P_c \rangle$ in the frontal region as
function of the relative height from the bottom of the front,
$(h-h_{\text{min}})/w_s$. The height is normalized by dividing with
the saturated front width $w_s$. In the simulations $\langle P_c
\rangle$ was computed by taking the average of the capillary pressures
of the front menisci at same height, $h$, above the inlet.
\fig{pcgrad} shows the result for two simulations with almost equal
$C_a$ but different $M$. One with $M=1$ and $C_a=\exf{1.6}{-3}$
(\tbl{m1}) and the other with $M=100$ and $C_a=\exf{1.0}{-3}$
(\tbl{m100-2}). If we consider the middle part of the front between
the two vertical dashed lines in \fig{pcgrad}, we observe that the
capillary pressure drop, $-w_sd\langle P_c\rangle/dh$, over a length
$w_s$ in the front, is higher for $M=100$ than for $M=1$, even though
the capillary numbers are almost equal. In both simulations a typical
narrow front with a compact displacement structure developed.  On
average, $-w_sd\langle P_c\rangle/dh$ must equal the difference
between the pressure drops taken in the nonwetting and wetting part of
the front over a length $w_s$ (see \fig{model}). When the nonwetting
and wetting fluid have equal viscosities the pressure drops in the
nonwetting and wetting part of the front is about the same, explaining
the smaller capillary pressure drop when $M=1$ than when $M=100$ in
\fig{pcgrad}.

\begin{figure}
\begin{center}
\mbox{\psfig{figure=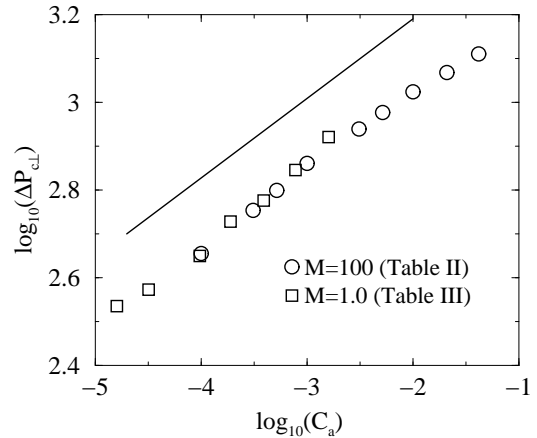,width=7cm}}
\caption{The logarithm of the plateau of $\Delta P_{c\bot}$ versus the
logarithm of $C_a$ for $M=100$ (circles) and $M=1$ (boxes)
corresponding to simulations listed in \protect\tbl{m100-2} and
\protect\tbl{m1}, respectively.  The slope of the solid line is
$0.2$. See also the inset of \protect\fig{xpos-pch}.}
\label{fig:pch-aver-log}
\end{center}
\end{figure}

Let us now study the behavior of $\Delta P_{c\bot}$. Simulations show
that $\Delta P_{c\bot}$ as function of $\Delta l$ does not change much
when comparing simulations performed at equal $C_a$ with $M=1$ and
$M=100$.  Especially, the constant plateau where the capillary
pressures are uncorrelated (see inset of \fig{xpos-pch}), has the same
value. This is illustrated in \fig{pch-aver-log} where we have plotted
the plateau of $\Delta P_{c\bot}$ versus $C_a$ in a logarithmic plot
for simulations with $M=100$ (\tbl{m100-2}) and $M=1$ (\tbl{m1}). From
the figure we observe that the plateau does not depend on $M$. As a
side mark, we notice that there seems to be a power law between the
plateau of $\Delta P_{c\bot}$ and $C_a$, which we indicate by the
straight line in \fig{pch-aver-log}.  The slope of the line is 0.2.

From the above discussion we draw the following conclusion.  Consider
two parallel and horizontal lines intersecting the front and let the
lines be separated a vertical distance $w_s$. When $M=1$ we have found
that the capillary pressure drop between the lines is small due to the
equal fluid viscosities (\fig{pcgrad}).  However, the magnitude
(plateau) of $\Delta P_{c\bot}$, is found to be the same as when
$M=100$ (\fig{pch-aver-log}). Thus, when $M=1$ the relative small
capillary pressure drop is annihilated by the magnitude of the
capillary variations in the horizontal direction, $\Delta P_{c\bot}$.
This destroys a possible power law behavior of $\Delta P_{c\|}(w_s)$ when
$M=1$ in \fig{pcdmax}. When $M=100$, the capillary variations are too
small to annihilate the larger capillary pressure drop there, giving
the increasing function $\Delta P_{c\|}(w_s)$. If we divide the
capillary pressure drop, calculated in \fig{pcgrad}, with the plateau
of $\Delta P_{c\bot}$ in \fig{pch-aver-log}, we find that the ratio is
a factor three lower for $M=1$ than for $M=100$ at
$C_a\simeq\exf{1.0}{-3}$.

\subsection{Capillary pressure on IP patterns\label{sec:ip}}
We have studied the capillary pressure along the front of patterns
generated by an IP algorithm with a stabilizing
gradient. The patterns were loaded into our network model, and the
simulations were started from that point.  Using this
method, we were able to perform displacement simulations in a short period of
time on patterns generated on lattices of $200\times 300$ nodes. The
result of these simulations are based on the assumption that the
generated patterns are statistically equal to the structures that
would have been obtained in a corresponding complete displacement
simulation.

The IP algorithm was performed on the bonds in a square lattice with
the bonds oriented at $45^\circ$. Hence, the bonds correspond to the
tubes in our network model and an occupied bond refers to a tube
filled with nonwetting fluid.  Each bond were assign a random number
$f_{ij}$ in the interval $[0,1]$ where $ij$ denote the bond between
the $i$th and the $j$th node in the lattice.  A stabilizing gradient
$g$ was applied on the lattice giving an occupation threshold $t_{ij}$
of every bond like, $t_{ij}=f_{ij}+gh_{ij}$~\cite{Wilk86,Birov91}.
Here $h_{ij}$ denotes the height of bond $ij$ above the bottom row.
The occupation of bonds started at the bottom row, and new bonds were
occupied until the invasion front reached the top row. There was
periodic boundary conditions in the horizontal direction.  The next
bond to be occupied was defined as the bond with the lowest threshold
value from the set of empty bonds along the invasion front. The
invasion front was found by running a Hoshen-Kopelman algorithm on the
lattice.

\begin{figure}
\begin{center}
\mbox{\psfig{figure=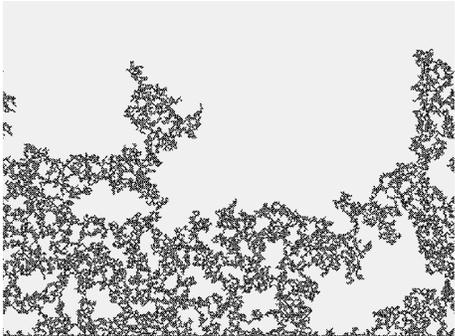,width=6cm}}
\caption{One of the generated IP patterns with $g=0.05$ on a lattice
of $200\times 300$ nodes. The pattern was loaded into our network
model.}
\label{fig:ip-structure}
\end{center}
\end{figure}

We generated four IP patterns with $g=0.05$ and different sets of
random numbers $f_{ij}$.  When the invasion front became well
developed with trapped (wetting) clusters of all sizes between the
size of the bonds and the front width, the structures were loaded into
our network model. \fig{ip-structure} shows one of the generated
IP patterns.

The loading was performed by filling the tubes in the network model
with nonwetting and wetting fluid according to occupied and empty
bonds in the IP lattice.  Furthermore, the radii $r_{ij}$ of the tubes
were mapped to the random numbers $f_{ij}$ of the bonds like,
$r_{ij}=[\lambda_1+(\lambda_2-\lambda_1)(1-f_{ij})]d$. Thus,
$r_{ij}\in [\lambda_1 d,\lambda_2 d]$ where, we set the tube length
$d=1\ \mbox{mm}$, $\lambda_1=0.05$, and $\lambda_2=1.0$.

Above, $r_{ij}$ is mapped to $1-f_{ij}$ because in the IP algorithm
the next bond to be invaded is the one with the lowest threshold
value, opposite to the network model, where the widest tubes will be
invaded first. Note also, that in the network model the invasion of
nonwetting fluid is controlled by the threshold capillary pressures
$p_{t}$ of the tubes. According to \eq{pcvary}
$p_t=4\gamma/r$ in the middle of the tubes where $x=d/2$. In the
IP model the distribution of $f_{ij}$ is flat. Thus, when 
$r_{ij}$ is mapped to $f_{ij}$ as described above, we obtain
a $1/{p_t}^2$ distribution of capillary pressure thresholds. However,
since there is a one to one correspondence in the mapping between
$f_{ij}$ and $p_t$, we can assume that the IP patterns are
statistically equal to similar structures that would have been
generated in a full displacement simulation. The assumption provides
that the displacement simulation is performed with an appropriate
injection rate $Q$, according to $g$ that was used to generate the IP
patterns.

After the IP patterns were successfully loaded into the network
model, we started the simulations and ran the displacement a limited
number of time steps while $\Delta P_{c\|}$ was recorded.  The number
of time steps were chosen such that the front menisci got sufficient
time to adjust according to the viscous pressure set up by the
injection rate. For all four structures we chose $M=100$ and $Q=0.1\
\mbox{ml/min}$, giving $C_a=\exf{9.5}{-5}$. This $C_a$ might
be too high compared to the front widths we obtained at low $C_a$ from
simulations listed in Tables~\ref{tbl:m100-1}
and~\ref{tbl:m100-2}. The reason why we choose a high $C_a$ is to
minimize computation time.  Simulations show that fewer time steps and
hence, less CPU time are required to adjust the front menisci when a
high injection rate is applied instead of a low one.  Moreover, the
simulations also show that as long as the number of time steps are
chosen sufficiently large to allow the front menisci to adjust, the
exponent $\kappa$ in $\Delta P_{c\|}\propto\Delta h^\kappa$, is not
sensitive on the injection rate. In the present simulations the number
of time steps was 400.

\begin{figure}
\begin{center}
\mbox{\psfig{figure=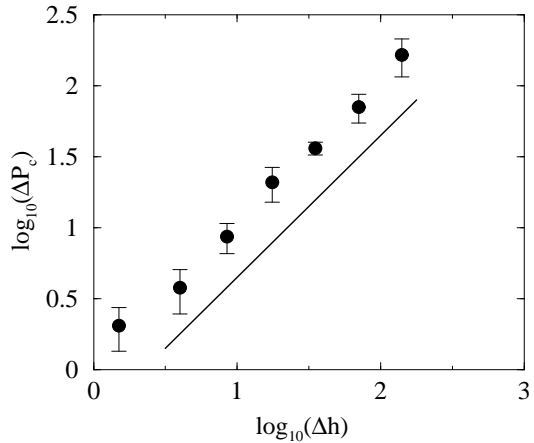,width=7cm}}
\caption{$\log_{10}(\Delta P_{c\|})$ as function of $\log_{10}(\Delta
h)$ for simulations initiated on IP patterns on lattices of
$200\times 300$ nodes. $C_a=\protect\exf{9.5}{-5}$ and $M=100$. The
result is averaged over four different runs and the error bars denote
the standard error in the mean. The slope of the straight solid line
is 1.0.}
\label{fig:pcd-ip}
\end{center}
\end{figure}

The result of the simulations is shown in \fig{pcd-ip} where we have
plotted $\log_{10}(\Delta P_{c\|})$ versus $\log_{10}(\Delta h)$. As
for the previous results, we find $\kappa = 1.0\pm 0.1$. The slope of
the straight line in \fig{pcd-ip} is 1.0. We have also done
displacement simulations on one of the IP patterns at 
$C_a=\exf{2}{-6}$ with $M=1$ and $M=100$. These simulations
were run in 1600 time steps and the result of those is consistent with
\fig{pcd-ip}.

\section{Effect of loopless strands}\label{sec:trap}

In~\cite{Xu98} it was argued that $\Delta P_{c\|}=\Delta P_{nw}-\Delta
P_w$ (see \fig{model}).  At low injection rates or when the nonwetting
phase is much more viscous than the wetting phase, $\Delta
P_{w}\ll\Delta P_{nw}$, giving $\Delta P_{c\|}\sim \Delta P_{nw}$.
Thus, if the result of Xu {\em et al.}~\cite{Xu98} should be valid for
our problem, we would expect to find $\Delta P_{c\|}\propto\Delta
h^\kappa$ where $\kappa=t/\nu+d_{\text{E}}-1-\beta/\nu$. Inserting
values of the exponents in 2D ($t=1.3$, $\nu=4/3$, $d_{\text{E}}=2$,
$\beta=5/36$) gives $\kappa\simeq 1.9$.  Our simulations clearly
indicate that $\kappa\simeq 1.0$ which is inconsistent with the
proposed result in~\cite{Xu98}.  Below we present an alternative view
on the displacement pattern from that being initiated by
Wilkinson~\cite{Wilk86} and used by Xu {\em et al.}.  The alternative
view is based upon the observation that nonwetting fluid flows in
separate strands.

\begin{figure}
\begin{center}
\setlength{\unitlength}{1cm}
\begin{picture}(8.5,12)
\put(0.7,6.2){\psfig{figure=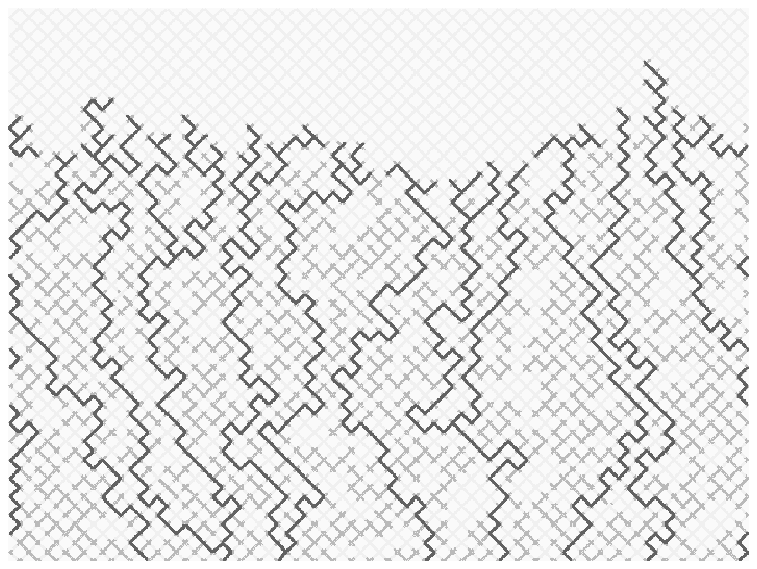,width=7cm}}
\put(2.5,5.9){\mbox{\footnotesize $C_a=3.9 \times 10^{-4},\ M=1.0$}}
\put(0.7,0.3){\psfig{figure=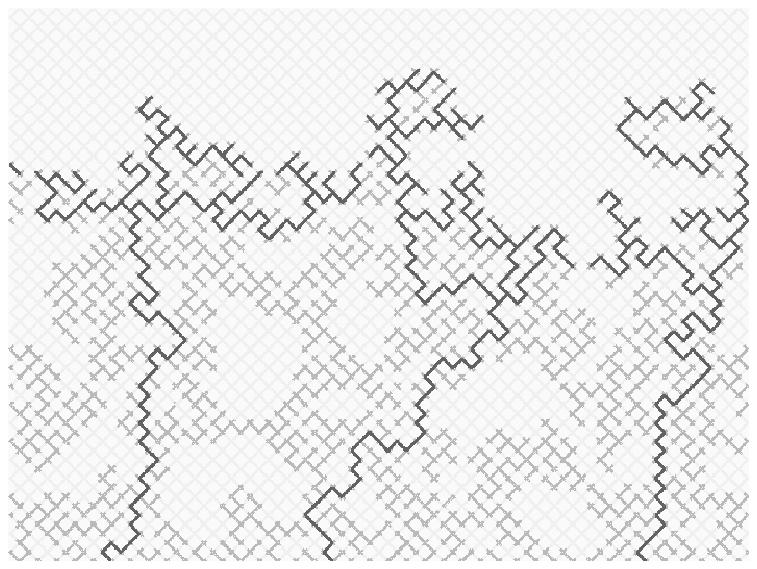,width=7cm}}
\put(2.5,0.0){\mbox{\footnotesize $C_a=1.6 \times 10^{-5},\ M=1.0$}}
\end{picture}
\caption{Two displacement structures of simulations at high
$C_a=\exf{3.9}{-4}$ (above) and low $C_a=\exf{1.6}{-5}$ (below) before
breakthrough of nonwetting fluid. The nonwetting fluid (dark grey and
black) is injected from below and wetting fluid (light grey) flows out
along the top row.  The lattice size was $40\times 60$ nodes and $M=1$
(Table~\protect\ref{tbl:m1}).  The black tubes denote the loopless
strands where nonwetting fluid flows and the dark grey tubes indicate
nonwetting fluid unable to flow due to trapped regions of wetting
fluid. Because of fluid trapping and subsequent volume conservation of
wetting fluid, strands of different starting points along the inlet
can never connect. Note the few fluid supplying strands from the inlet
to the frontal region at low $C_a$ compared to the case at high
$C_a$.}
\label{fig:40x60}
\end{center}
\end{figure}
\fig{40x60} shows two typical displacement structures that were
obtained from simulations at low and high $C_a$ on the lattice of
$40\times 60$ nodes with $M=1$ (\tbl{m1}). We observe that the
nonwetting fluid (dark grey and black) generates patterns containing
no closed loops. That means, following a path on nonwetting fluid will
never bring us back to the starting point. The loopless structure is a
direct consequence of the evidence that a tube filled with wetting
fluid and surrounded on both sides by nonwetting fluid is trapped due
to volume conservation of wetting fluid. Because of trapped wetting
fluid, the nonwetting fluid also flows in separate strands, indicated
as black tubes in \fig{40x60}.  When the nonwetting fluid percolates
the system there exists only on such strand connecting the inlet to
the outlet.  The dark grey tubes connecting to the strands are dead
ends where nonwetting fluid cannot flow because of trapped wetting
fluid. We note that the evidence of trapped wetting fluid in single
tubes may easily be generalized to 3D and therefore our arguments
should be valid there too.  Similar loopless structures as in
\fig{40x60}, were also pointed out in~\cite{Sahimi98} for site-bond IP
with trapping and in~\cite{Cieplak96} for a loopless IP algorithm.

From \fig{40x60} we may separate the displacement patterns into two
parts. One consisting of the frontal region continuously covering new
tubes, and the other consisting of the more static structure behind
the front. The frontal region is supplied by nonwetting fluid through
a set of strands that connect the frontal region to the inlet. When
the strands approach the frontal region they are more likely to
split. Since we are dealing with a square lattice, a splitting strand
may create either two or three new strands.  As the strands proceed
upwards in \fig{40x60}, repeatedly splits cause the frontal region
to be completely covered by nonwetting strands.

On IP patterns with trapping~\cite{Sahimi98} or without
loops~\cite{Cieplak96,Porto97} the length $l$ of the minimum path
between two points separated an Euclidean distance $R$ scales like
$l\propto R^{D_s}$ where $D_s$ is the fractal dimension of the
shortest path. We assume that the displacement pattern of the frontal
region for length less than the correlation length (in our case $w_s$)
is statistically equal to IP patterns in~\cite{Sahimi98}.  Therefore,
the length of the nonwetting strands in the frontal region, is
proportional to $\Delta h^{D_s}$ where $\Delta h$ is some vertical
length less than $w_s$. If we assume that on the average every tube in
the lattice has same mobility ($k_{ij}/\mu_{ij}$), we obtain that the
fluid pressure within one strand must drop like $\Delta h^\kappa$
where $\kappa=D_s$. Let us now consider the effect on the pressure
when strands split. If we assume that the strands are straight
($D_s=1$) then following a path where strands splits would cause the
pressure to drop as $\Delta h^\kappa$ where $\kappa < 1$. This because
the volume fluxes through the strands after a split must be less than
the flux in the strand before it splits, due to volume conservation of
nonwetting fluid.

The two effects ($\kappa=D_s$ and $\kappa<1$) predict that the
pressure drop in the nonwetting phase of the frontal region, $\Delta
P_{nw}$, should scale as $\Delta P_{nw}\propto\Delta h^\kappa$ where
$\kappa\le D_s$. In 2D two different values for $D_s$ have been
reported: $D_s=1.22$~\cite{Cieplak96,Porto97} for loopless IP
patterns, and $D_s=1.14$~\cite{Sahimi98} for the single strand
connecting the inlet to the outlet when nonwetting fluid percolates
the system. We note that the result in~\cite{Sahimi98} is essential
equal to $D_{\text{min}}= 1.13$~\cite{Stauf92}, that is the fractal
dimension of the minimum path in 2D percolation where loops generally
occur.  Any of the above values for $D_s$ together with the argument
$\kappa\le D_s$, are supported by our simulations finding $\kappa =
1.0\pm 0.1$.

Note the different pattern of strands at high and low $C_a$ in
\fig{40x60}.  At low $C_a$ few strands are supplying the frontal
region with nonwetting fluid, and the strands split many times before
the whole front is covered.  At high $C_a$ the horizontal distance
between each strand in the static structure is much shorter, and only
a few splits are required to cover the front. Moreover, we observe
that at high $C_a$ the length of individual strands in the front
approaches the minimum length due to the tubes. In this limit we may
treat the strands in the front as straight lines (i.e. $D_s=1$)
causing $\kappa\le 1$. This is indeed supported by numerical results,
finding that $\kappa$ decreases from about $1.0$ to $0.8$ when
increasing $C_a$ (see \fig{pcdiff}).

Another important issue, arising at low $C_a$, is the effect of bursts
on the capillary pressure. A burst occurs when a meniscus along the
front becomes unstable and nonwetting fluid abruptly covers new
tubes~\cite{Maloy96}. The strand where the burst initiates will during
the burst, experiences a much higher fluid transport relative to
strands far away. Describing the pressure behavior between the strand
of the burst and the rest of the front is nontrivial. However,
simulations show that even during bursts, we find that $\Delta
P_{c\|}$ increases linearly with $\Delta h$.

The indication that $\kappa\simeq 1.0$, may influence the scaling behavior
of $w_s$ as function of $C_a$. Assuming Darcy flow where the pressure
drop depends linearly on the injection rate, we conjecture that
$\Delta\widehat{P}_{c\|}\propto C_a\Delta h^{\kappa}$. Here
$\Delta\widehat{P}_{c\|}$ denotes the capillary pressure difference
over a height $\Delta h$ when the front is stationary. That means,
$\Delta\widehat{P}_{c\|}$ excludes situations where nonwetting fluid
rapidly invades new tubes due to local instabilities (i.e. bursts).
The above conjecture is supported by simulations showing that in the
low $C_a$ regime $\Delta\widehat{P}_{c\|}\propto C_a\Delta
h^{\kappa}$ where $\kappa\simeq 1.0$.  Note, that
$\Delta\widehat{P}_{c\|}\not\simeq \Delta P_{c\|}$ in \fig{pcdiff},
since the latter includes both stable situations and bursts.

At sufficiently low $C_a$ the displacement may be mapped to
percolation giving $\Delta\widehat{P}_{c\|}\propto
f-f_c\propto\xi^{-1/\nu}$~\cite{Sap88,Wilk86,Birov91}. Here $f$ is the
occupation probability of the bonds, $f_c$ is the percolation
threshold, and $\xi\propto w_s$ is the correlation length. By
combining the above relations for $\Delta\widehat{P}_{c\|}$ we obtain
$w_s\propto {C_a}^{-\alpha}$ where $\alpha=\nu/(1+\nu\kappa)$. In 2D
$\nu =4/3$ and inserting $\kappa=1.0$ gives $\alpha\simeq 0.57$.

In Sec.~\ref{sec:pc} we found that at high $C_a$ the nonwetting fluid
invades simultaneously everywhere along the front. Hence, the front
never reaches a stationary state because of rapidly succeeding local
instabilities. This is supported by simulations showing a crossover in
$\Delta\widehat{P}_{c\|}$ to a nonlinear dependency on
$C_a$. Consequently, the above mapping to percolation might no longer
be valid and we expect another type of functional behavior between
$w_s$ and $C_a$ in the high $C_a$ regime.

\section{Comparison with Experiments}\label{sec:expr}

\begin{figure}
\begin{center}
\mbox{\psfig{figure=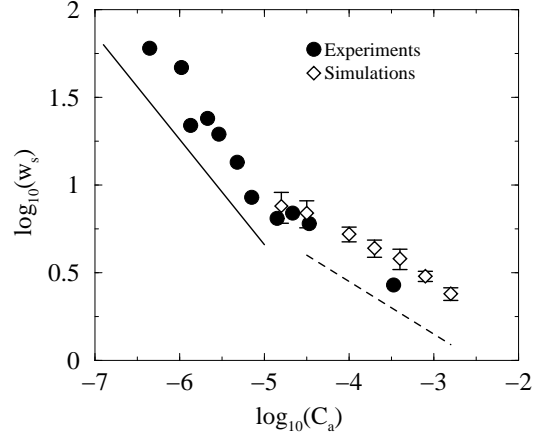,width=7cm}}
\caption{$\log_{10}(w_s)$ as function of $\log_{10}(C_a)$ for
experiments from~\protect\cite{Frette97} and simulations on the
lattice of $40\times 60$ nodes (Table~\protect\ref{tbl:m1}). For both
experiments and simulations $M=1$. The slope of the solid and dashed
line is -0.6 and -0.3, respectively.}
\label{fig:w-Ca}
\end{center}
\end{figure}

Frette {\em et al.}\ \cite{Frette97} performed two phase drainage
displacement experiments in a 2D porous medium with viscosity matched
fluids ($M=1$).  They reported on the stabilization of the front and
measured the saturated front width $w_s$, as function of $C_a$. For
all our simulations except those performed on the IP patterns, we
have calculated $w_s$. In \fig{w-Ca} we have plotted $w_s$
as function of $C_a$ in a logarithmic plot for the simulations in \tbl{m1},
(open diamonds) together with the experimental data of Frette {\em et al.}
(filled circles).

In~\cite{Frette97}, their best estimate of the exponent $\alpha$ when
assuming a power law $w_s\propto {C_a}^{-\alpha}$ was $\alpha=0.6\pm
0.2$, indicated by the solid line in \fig{w-Ca}.  This is consistent
with the suggested value $\alpha = \nu/(1+\nu\kappa)\simeq
0.57$ from Sec.~\ref{sec:trap}. The simulations show a different
behavior and they seem to fit $\alpha=0.3\pm 0.1$, according to the
dashed line in \fig{w-Ca}. The simulations performed on the lattices
of $25\times 35$ nodes (Tables~\ref{tbl:m100-1} and~\ref{tbl:m100-2})
also give $\alpha\simeq 0.3$.

Even though the overlap between experimental and numerical data in
\fig{w-Ca} is poor we suggest that the different behavior of the
experiments (at $C_a\lesssim\exf{1.0}{-5}$) and simulations (at
$C_a\gtrsim\exf{1.0}{-5}$) might be due to an expected change in
$\alpha$ at high $C_a$. According to the discussion in
Sec.~\ref{sec:trap} it is not clear if the percolation approach giving
$\alpha=\nu/(1+\nu\kappa)$, is valid for high $C_a$. The different
scaling behavior observed in \fig{w-Ca} might also be caused by the
small system size of the simulations. At $C_a\simeq\exf{1.0}{-5}$
numerical simulations show that the front width becomes bounded by the
system size, and therefore we are not able to observe a possible
$\alpha=\nu/(1+\nu\kappa)$ regime. We stress that more simulations on
larger systems and at lower $C_a$ are required in order obtain better
overlap between simulations and experiments in~\fig{w-Ca}. Until then,
it is hard to draw any conclusions on the correct $\alpha$.

As a side mark, we note that our simulations giving $\alpha\simeq 0.3$, are
in agreement with numerical work in~\cite{Xu98}. Their calculations of
$w_s$ were done for $C_a$ between $10^{-5}$ and $10^{-4}$ coinciding
with our region of simulations in~\fig{w-Ca}. According to
Wilkinson~\cite{Wilk86} $\alpha = \nu/(1+t-\beta+\nu)$ and by
inserting values of the exponents in 2D we obtain $\alpha\simeq
0.38$. This is also within the uncertainties of our simulation
results.  However, we emphasize that this might as well be a
coincidence rather than an evidence, because Wilkinson's theory does
not take into account that nonwetting fluid flows in strands along the 
front.

A somewhat different process, but very interesting result, is presented by
Shaw in~\cite{Shaw87}. He measured the width of the drying front in a
quasi 2D porous system and found that $w_s\propto {v_f}^{0.48\pm
0.1}$. Here $v_f$ is the average front velocity. Quite recently, this
has been compared to theory in~\cite{Yortsos99}.

\section{Conclusion}\label{sec:concl}

We have reported on the stabilization mechanisms of the front in
drainage displacement going from low to high injection rates.  The
stabilization process was studied by using a network model simulating
the viscous and capillary pressure buildup in the fluids during the
displacements. We have found that the capillary pressure difference
$\Delta P_{c\|}$, along the front varies almost linearly with the
distance $\Delta h$, in the direction of the displacement. We conclude
from simulations that $\Delta P_{c\|}\propto \Delta h^\kappa$ where
our best estimate is $\kappa=1.0\pm 0.1$.  This result supports the
arguments showing $\kappa\le D_s$, where $D_s$ is the fractal
dimension of the loopless strands characterizing the displacement
pattern. The evidence that nonwetting fluid flows in loopless strands
along the front are not considered in earlier proposed
theories~\cite{Wilk86,Len89,Blunt92,Xu98}. Hence, we conclude that they are
not compatible with drainage when nonwetting strands dominate the
displacement process.

Using the evidence that $\kappa\simeq 1.0$, we conjecture that the scaling
of the front width $w_s$ as function of $C_a$ might alters from
earlier suggestions in~\cite{Wilk86,Blunt92,Xu98}. By mapping our
problem to percolation we find $w_s\propto{C_a}^{-\alpha}$ where
$\alpha=\nu/(1+\nu\kappa)$. The result is consistent with experiments
performed by Frette {\em et al.}~\cite{Frette97}.  Unfortunately, due
to the small system sizes we are not able to confirm this scaling
behavior by our simulations.  We emphasize that a more stringent test
on $\alpha$ should include simulations on larger systems and lower
$C_a$, than presented here.

In addition to $\Delta P_{c\|}$ we have calculated the capillary
pressure variations along the front in the direction parallel to the
inlet, $\Delta P_{c\bot}$. Qualitatively, we have shown that $\Delta
P_{c\bot}$ is a good indicator on whether the capillary pressures of
the menisci along the front are all equal (capillary equilibrium) or
fluctuating due to the viscous forces.  When the capillary
fluctuations are strong, we do not expect percolation to be a proper
model for the displacement process.

\acknowledgments

The authors thank J.\ Feder, E.\ G.\ Flekk\o y for valuable comments.
The work is supported by the Norwegian Research Council (NFR) through a
``SUP'' program and we acknowledge them for a grant of computer time.

\appendix

\section{}\label{app:alex}

Below we show how to deduce $\alpha=\nu/(1+t-\beta+\nu)$ in
$w_s\propto {C_a}^{-\alpha}$ and find the corresponding exponent
$\kappa=1+t/\nu+\beta/\nu$ in the power law $\Delta P_{c\|}\propto
\Delta h^\kappa$ when not considering that nonwetting fluid flows
through strands. The calculations are carried out in two dimension,
however the extension to three dimensions is straight forward.

Let us consider a piece of the nonwetting phase of size $\Delta h$ in
the frontal region. We assume that $\Delta P_{c\|}$ vary as
\begin{equation}
\Delta P_{c\|}\propto v\Delta h^\kappa,
\label{eq:Pc}
\end{equation}
where $v$ is the average fluid velocity in the pores.  Moreover, we
assume that the front has reached a steady state and that the
structure of the front is statistically equal to the front of an
invasion percolation pattern.  This assumption provides that $\Delta
h$ is sufficiently large for the percolation concept to apply but less
than the front width $w_s$.

The average nonwetting pore fluid velocity $v$, in the 
the region of size $\Delta h$, is given by Darcy's law 
\begin{equation}
v=\frac{1}{S}\frac{k}{\mu}\frac{\Delta P_{c\|}}{\Delta h}.
\label{eq:v}
\end{equation}
Here $S$ is the saturation of nonwetting phase, that is the volume
fraction where nonwetting fluid can flow, and $k$ is the permeability
of the frontal region. According to percolation the frontal region is
fractal, with fractal dimension $D=d-\beta/\nu$, giving
\begin{equation}
S\propto\frac{\Delta h^{d-\beta/\nu}}{\Delta h^d}=\Delta h^{-\beta/\nu},
\label{eq:S}
\end{equation}
and
\begin{equation}
k\propto \Delta h^{-t/\nu}.
\label{eq:k}
\end{equation}
Here $t$ is the conductivity exponent, $\beta$ is the order
parameter exponent, and $\nu $ is the correlation length exponent
in percolation.

By inserting the expressions for $S$, $k$, and $\Delta P_{c\|}$ into
\eq{v} we find the exponent $\kappa = 1 + t/\nu - \beta/\nu$.  The
exponent $\alpha$ follows by setting $\Delta h=w_s$ and replace
$\Delta P_{c\|}$ in \eq{Pc} with the power law $w_s\propto\xi\propto
\Delta P_{c\|}^{-\nu}$. Here $\xi$ denote the correlation length in
percolation.

\end{document}